\documentstyle[preprint,aps]{revtex}
\begin{document}
\preprint{\vbox{\rightline{IFT-P.051/99}}}
\draft
\title{Reply to the comment by D. Kreimer and E. Mielke.}
\author{Osvaldo Chand\'{\i}a$^a$ and Jorge Zanelli$^{b,c}$}
\maketitle
\begin{center}
$^{(a)}$ Instituto de F\'{\i}sica Te\'orica, Universidade Estadual Paulista, \\
Rua Pamplona 145, S\~ao Paulo, Brasil\\
$^{(b)}$ Centro de Estudios Cient\'{\i}ficos de Santiago, Casilla 16443,
Santiago, Chile\\ $^{(c)}$ Departamento de F\'{\i}sica, Universidad de
Santiago de Chile, Casilla 307, Santiago 2, Chile 
\end{center}

\begin{abstract}
We respond to the comment by Kreimer et. al. about the torsional
contribution to chiral anomaly in curved spacetimes. We discuss their
claims and refute its main conclusion.
\end{abstract}

In the article by D. Kreimer and E. W. Mielke \cite{K-M} the existence of a
chiral anomaly in spaces with torsion as reported by us in \cite{Ch-Z1} is
challenged. Their result is presumably based on standard diagramatic
techniques and regularization prescriptions. Our results, on the other hand,
are derived using functional methods to evaluate formally the expectation
value of the divergence of the chiral current as the regularized trace of the
chiral operator.

Their claim that the Nieh-Yan four-form has been shown to be irrelevant for
the anomaly is contradictory with subsequent confirmations by Obukhov et al.
(second article in \cite{obukhov}) who obtain the same anomaly in the heat
kernel approach, Soo \cite{Soo} and Chang and Soo \cite{C-S}, who apply the
Pauli-Villars technique in a standard diagramatic approach, and our own
\cite{Ch-Z2,Ch-Z3}, where the anomaly is obtained as the Atiyah-Singer index
using supersymmetry. So, apart from the puzzling fact that one of the authors
of the comment is also coauthor of the paper by Obukhov et al., it is not at
all clear that our result was demonstratedly incorrect.

In our view there are a number of unproven and false statements in the
comment which we would like to point out. As a result, the main claim of the
comment cannot be substantiated. Below we analyze those statements one by one
as they appear in the text.

1)(Section I, second paragraph) ``The fact that the anomaly is stable
against radiative corrections guarantees that it can be given a topological
interpretation.'' 

{\bf Response:} Here the order should be reversed: the anomaly is stable
under perturbative corrections {\em because} it is a topological invariant.

2)(Third paragraph) ``...there is no doubt that the NY term can be possibly
generated, as demonstrated previously \cite{obukhov,Y},..'' 

{\bf Response:} It should be clarified that the second article in Ref.
\cite{obukhov} did not precede ours. As stated in the abstract of Ref.
\cite{obukhov}, ``Following Chandia and Zanelli, two spaces with non-trivial
translational Chern-Simons forms are  discussed. We then demonstrate, firstly
within the classical Einstein-Cartan-Dirac theory and secondly in the quantum
heat kernel approach to the Dirac operator, how the Nieh-Yan form surfaces in
both contexts, in contrast to what has been assumed previously.''. Moreover,
although the NY form can be found in the first article of Ref.\cite{obukhov}
and in Ref.\cite{Y}, it is not identified as a topological invariant and mixed
with dozens of other terms which are collectively discarded as irrelevant. So,
to say that ``in Refs. \cite{obukhov,Y} it was demonstrated that the NY term
is undoubtedly generated'' is a gross overstatement. To set the record
straight, we were the first to point out that the topological NY-four form is
generated in Einstein-Cartan spaces.

3)(Third and fourth paragraphs) ``In rescaling the tetrad, the authors of
Ref. \cite{Ch-Z1} ignore the presence of renormalization conditions and the
generation of a scale upon renormalization. Rescaling the tetrad would
ultimately change the wave function renormalization Z-factor...This factor
creeps into the definition of the NY term at the quantum level, and thus a
rescaling of the tetrad does not achieve the desired goals... With no
renormalization condition available for the NY term, and other methods
obtaining it as zero, we can only conclude that [the anomaly] delivers no NY
term. Or, saying it differently, its finite value is zero after
renormalization.''

{\bf Response:} Anomaly calculations assume a given background --which is
not necessarily quantized-- and a quantized chiral field. Pertubative
radiative corrections cannot change the a topological density, which is a
function of the background field [Adler-Bardeen-Zumino theorem]. Then, it is
sufficient to check that the result is a topological invariant (including
scale invariance, of course), in order to be sure it does not renormalize.
Kreimer et al. do not show --in this paper or elsewhere-- that the  rescaled
Nieh-Yan four form, $l^{-2}[T^a T_a - R_{ab} e^a e^b]$ {\em is not} a
topological invariant, nor have they shown that renormalization does what they
claim. Hence, there is no ground to claim that our result is spurious and will
be erased by renormalization. 

4) (Section II, third paragraph) ``Since the coframe is the translational
part of the Cartan connection...''

{\bf Response:} That identification rests on the assumption that the tangent
space be flat and with a Poincar\'e fiber. Since Minkowski space has no scale,
identifying the vielbein with the ``connection'' associated to translations is
a slippery issue: the vierbein has dimensions of length, while the connection
is dimensionless. Hence, one is forced to introduce an artificial length scale
in a scale invariant space. Instead, we would prefer to identify the tangent
space with a manifold of constant curvature with (A)dS local invariance. This
has the advantage of having a naturally defined length scale $l$~(radius of
curvature).

5) (Fourth paragraph)``The corresponding CS term $\hat{C}$ spplits via
$\hat{C}=C_{RR} -2C_{TT}$ into the linear one and that of translations, see
the footnote 31 of Ref.\cite{Hehl}. This relation has recently been
``recovered'' by Chandia and Zanelli \cite{Ch-Z1}.''

{\bf Response:} Here the wording and the use of quotation marks is meant to
imply that we have appropriated an idea found in Ref. \cite{Hehl} without
acknowledging it. This splitting was ``recovered'' by the present authors as
much as by the authors of Ref. \cite{Hehl}. It is first mentioned in the
original papers by Nieh and Yan \cite{NY}, and further discussed also, several
years prior to Ref. \cite{Hehl}, in a paper by A. Mardones and J. Zanelli
\cite{M-Z}. What was not at all clear in the literature before our paper, was
the topological nature of the NY invariant or its relationship with the Chern
classes of SO(5) and SO(4), and much less, its relevance for the chiral
anomaly.

6) (Section III, second paragraph) ``The decomposed Lagrangian (3.1) leads
to the following form of the Dirac equation \[
i*\gamma \wedge \breve{D} \psi + *m\psi = i*\gamma \wedge\left[
D^{\{\}}+\frac{i}{4} m\gamma +\frac{i}{4}A\gamma_5 \right] \psi =0
\;\;\;\;\;\;\; (3.2) \] in terms of the Riemannian connection $\Gamma^{\{\}}$
with $D^{\{\}}\gamma =0$ and the {\em irreducible} piece (2.3) in the torsion.
Hence, in a RC spacetime a Dirac operator does only feel the {\em axial
torsion} one-form $A$. This can also be seen from  the identity (3.6.13) of
Ref. \cite{Hehl} which specializes here to the ``on shell'' commutation
relation \[ [\breve{D}, \breve{D}] = \Omega^{\{\}} +\frac{i}{4}\gamma_5 dA
-\frac{i}{8}m^2\sigma. \;\;\;\;\;\; (3.3)'' \] 

{\bf Response:} Although the ``on shell'' Dirac operator $\breve{D}$ is not
clearly defined, it is not true that because other components of the torsion
do not appear in the Dirac equation (3.2), they will not contribute to the
interaction. The expression  for the commutator is incorrect. The correct form
is found in the literature and can be easily checked to be  \[ [D,D]=
\frac{1}{4}
J^{ab}J^{cd}R_{ab\,cd}-J^{ab}e_a^{\mu}e_b^{\nu}e_c^{\lambda}T^c_{\mu \nu}
D_{\lambda}. \] Clearly, if one uses the ``on shell'' relations, one could
cancel some part of the second term in the RHS. The  problem is then how to
justify using ``on shell'' relations which don't go through in the quantum
regime. It is precisely this last term that gives rise to the Nieh-Yan form
through the Fujikawa method, which obviously could not be reproduced starting
from (3.3).

7) (Fifth paragraph) ``From Einstein's equations ... and the purely
algebraic {\em Cartan relation}... one finds \[
dj_5 \cong 4dC_{TT} = \frac{2}{l^2} \left( T^{\alpha}\wedge T_{\alpha}
+R_{\alpha \beta} \wedge\vartheta^{\alpha}\vartheta^{\beta} \right) \;\;\;\;
(3.4) \] which establishes a link to the NY four form, but only for the {\em
massive} fields.''

{\bf Response:} There are three remarkable points here: the first is that in
spite of the above relation, the authors continue to believe that ``in a RC
spacetime a Dirac operator does only feel the {\em axial torsion} one-form
$A$'' (see 6, above). How can this be if at the same time the violation of
the fermionic chiral current is due to the interaction with torsion through
the NY torsional term?. The second point is their use of the Einstein
equations. Why should the Einstein equations be at all relevant to this
problem? In fact, as is well known, the integrability condition for the Dirac
operator is precisely Einstein's quations (in particular, that is why local
supersymmetry requires gravity, and that is one way in which supergravity
arises). Thus, had they computed the integrability condition correctly, they
would have found Einstein's equations and the rest of the argument would
follow naturally. The third remarkable point is their claim that the result be
valid for massive fields only. This is puzzling because there is no mass
parameter in (3.4) and nothing seems to prevent taking the limit $m
\rightarrow 0$. This last observation shows that there must be something fishy
about their next claim: 

8) (Sixth paragraph)``in the limit $m \rightarrow 0$, we find within the
dynamical framework of ECD theory that the NY four-form tends to zero ``on
shell'', i.e. $dC_{TT} \cong (1/4)dj_5 \rightarrow 0$''. 

{\bf Response:} Again, the suspect ``on shell'' relations are invoked to
justify an otherwise irreproducible result, because the only relation that
links $dj_5 $ with the mass are ``on shell'' equations. 

9) (Section IV, second and third paragraphs) ``[in order to calculate the
anomaly, we] concentrate on the last term [$-\frac{1}{4}A\wedge
\bar{\psi}\gamma_5*\gamma \psi$] in the Lagrangian... this term can be
regarded as an {\em external} axial covector $ A$ ... coupled to the axial
current $j_5$ of the Dirac field in an {\em initially flat} spacetime. By
applying the result (11-225) of Itzykson and Zuber..., we find that only the
term $dA\wedge dA$ arises in the axial anomaly, but {\em not} the NY type
term $d*A \cong dC_{TT}$ as was recently claimed \cite{Ch-Z1}.''

{\bf Response:} The calculation in Itzykson and Zuber (I-Z) would be valid
for commutation relations of the form (3.3), but unfortunately, as we said in
6), this is not the case. Eq. (3.3) is valid only on shell, but that is
insufficient to apply the I-Z  result, especially because this is supposed to
be a quantum calculation. So it is not that the I-Z approach is wrong, it is
just not designed to handle a Dirac operator that satisfies a more complicated
relation such as that in RC spaces, so it should be  rederived in order to
apply it to this case, something Kreimer and Mielke didn't do.

For the purposes of our work, the $U(1)$ anomaly is completely standard and
cannot yield anything new which was not there already, say, in
electrodynamics. In particular, since $\pi_3[U(1)]$ is trivial
\cite{nakahara}, its presence can be gauged away.

10) (Fifth paragraph) ``Whereas in $n=4$ dimensions the Pontrjagin type term
$K_4$ is dimensionless, the term $K_2 \sim2l^2 dC_{TT}$ carries dimensions. It
can be consistently absorbed in a counterterm, and thus discarded from the
final result for the anomaly.''

{\bf Response:} Kreimer and Mielke provide no proof of this claim. They do
not exhibit the counterterm or the radiative corrections that can give rise to
it. Although they often refer to renormalization and counterterms, there is
not a single one-loop calculation to be found anywhere in the paper.

11) (Seventh paragraph)``In Ref. \cite{Ch-Z1} it is argued that such
contributions can be maintained by absorbing the divergent factor in a
rescaled coframe $\tilde{\vartheta}^{\alpha} := M \vartheta^{\alpha}$ and
propose to consider the Wigner-In\"{o}n\" {u} contraction $M \rightarrow
\infty$ in the de Sitter gauge approach [6], with $Ml$ fixed.''

{\bf Response:} This is not true. We did not propose that rescaling. All we
did was to observe that if one replaces $\tilde{\vartheta}^{\alpha}$ by
$(Ml)^{-1} \vartheta^{\alpha}$ [Eq. (30) in our paper], the result reads,

\begin{equation}
{\cal A}(x) = \frac{1}{8\pi^2} \left[ R^{ab}\mbox{\tiny $\wedge$}R_{ab} +
\frac{2}{l^2} (T^a{\mbox{\tiny $\wedge$}}T_a - R_{ab}{\mbox{\tiny
$\wedge$}}e^a{\mbox{\tiny $\wedge$}}e^b )\right].  \label{result}
\end{equation}

This, in the language of Kreimer and Mielke is equal to $dC_{RR}+dC_{TT}$,
which is just the Chern class for $SO(5)$.

12) (Ninth paragraph) ``1. As the difference (2.1) of two Pontrjagin
classes, the term $dC_{TT}$ is a topological invariant after all. Now, it is
actually {\em not} this term which appears as the torsion-dependent extra
contribution to the anomaly, but more precisely $-d*A = 2l^2 dC_{TT}$. Thus,
measuring its proportion in units of the topological invariant $dC_{TT}$, we
find that it vanishes when we consider the proposed limit $M\rightarrow
\infty$, keeping $Ml$ constant.''

{\bf Response:} The redefinition $\vartheta \rightarrow (Ml)^{-1}\vartheta$,
does not change the units (assuming $c$ and $\hbar$ are dimensionless). The
scaling properties of $C_{TT} =l^{-2} \vartheta^a \wedge T_a$ depend on how
$\vartheta$ and $l$ are supposed to scale. If the fields are properly
defined, the one-form $\vartheta$ has dimensions of length and therefore
$l^{-1} \vartheta$ is dimensionless, like any well defined connection
one-form. Then, $C_{TT}$, $dC_{TT}$ and the anomaly are scale invariant as
they should.

13) (Tenth and eleventh paragraphs) ``...consistently a renormalization
condition can be imposed which guarantees the anomaly to have the
[torsion-free] value. Even if one renders this extra term finite by a
rescaling as in Ref. \cite{Ch-Z1}, one has to confront the fact that a
(finite) renormalization condition can be imposed which settles the anomaly at
this value.... From a {\em renormalization group} point of view, it is the
scaling of the coupling which determines the scaling of the anomaly... a
property which is [needed to satisfy the conditions of] the Adler-Bardeen
theorem. Or, to put it otherwise, an anomaly is stable against radiative
corrections for the reason that such corrections are compensated by a
renormalization of the coupling. While, on  the other hand, the topological
invariant of Ref. \cite{Ch-Z1} has no such property, its interpretation as an
anomaly seems dubious to us.''

{\bf Response:} As it is easily seen, Kreimer and Mielke's comment is more
of a warning about the problems one might encounter than a proof that
something wrong has actually been done. As they produce no evidence in support
of their contention, they end up in a sceptical remark. This is the most
honest claim in the entire comment (even if it is incorrect because the
topological invariant of Ref. \cite{Ch-Z1} does possess the property they
would like it to have).  In our opinion, they should have limited themselves
to just that last line.

14) (Fifth section) Here Kreimer and Mielke essentially repeat their claims
without adding any new arguments.

In conclusion, we can summarize the following points:

{\bf A} The authors of the comment do not argue against the fact that the NY
term is present in the chiral anomaly as we had shown. 

{\bf B} They challenge the contribution of the Nieh-Yan topological
invariant, on the grounds that they do not find it through manipulations in
which they use ``on shell'' conditions.

{\bf C} They furthermore claim that radiative corrections will renormalize
the NY term to zero, although they do not produce any evidence for this (e.g.,
loop corrections to the effective action, etc.). Their claim rests on a
scaling argument, according to which the NY term scales with the mass, which
is incorrect or at best arbitrary.

{\bf D} They criticize a rescaling argument which they attribute to us but
which is nowhere to be found in our paper.

\begin{center}
{\bf ACKNOWLEDGMENTS}
\end{center}

This work was partially supported by grants 1970151, 1980788 and 1990189 from
FONDECYT-Chile, grant 98/02380-3 from FAPESP-Brasil and grant 27-9531ZI from
DICYT-USACH. The institutional support by Fuerza A\'{e}rea de Chile and a
group of Chilean companies (AFP Provida, Business Design Associates, CGE,
CODELCO, COPEC, Empresas CMPC, GENER S.A., Minera Collahuasi, Minera
Escondida, NOVAGAS and XEROX-Chile) is also recognized.


\end{document}